\documentclass[10pt]{article}

\usepackage{amsmath}
\usepackage{amssymb}

\usepackage{graphicx}

\usepackage{cite}

\usepackage[pdftex]{color} 
\usepackage{hyperref} 

\usepackage[labelfont=bf,labelsep=period,justification=raggedright]{caption}

\makeatletter
\renewcommand{\@biblabel}[1]{\quad#1.}
\makeatother
\date{}

\newcommand{\kBT}{k_\text{B}T} 
\newcommand{\mean}[1]{\langle #1 \rangle} 
\newcommand*{\Var}{\text{Var}}

\newcommand{\Eq}[1]{Eq.~\eqref{#1}}

\newcommand{\mum}{\mu\text{m}}

\newcommand{\fix}[1]{\left. #1\right|}

\newcommand{\fixdiff}[2]{\fix{\frac{\partial #1}{\partial #2}}}

\newcommand{\nm}{\text{nm}}

\newcommand{\ML}[1]{#1}

\begin{document}

\begin{flushleft}
{\Large
\textbf{Entropic Tension in Crowded Membranes}
}
\\
Martin Lind\'en$^{1,2}$, 
Pierre Sens$^{3}$, 
Rob Phillips$^{1,3,4,\ast}$
\\
\bf{1} Dept. of Applied Physics, California Institute of Technology, Pasadena, California, U.S.A
\\
\bf{2} Present address: Dept. of Biochemistry and Biophysics, Stockholm University, Stockholm, Sweden
\\
\bf{3} Laboratoire de Physico-Chimie Th{\'e}orique CNRS/UMR 7083 - ESPCI, 75231 Paris Cedex 05, France

\bf{4} Division of Biology, California Institute of Technology, Pasadena, California, U.S.A
\\
$\ast$ E-mail: phillips@pboc.caltech.edu
\end{flushleft}

\pdfbookmark[1]{Abstract}{Abstract}
\section*{Abstract}
Unlike their model membrane counterparts, biological membranes are
richly decorated with a heterogeneous assembly of membrane proteins.
These proteins are so tightly packed that their excluded area
interactions can alter the free energy landscape controlling the
conformational transitions suffered by such proteins.  For membrane
channels, this effect can alter the critical membrane tension at which
they undergo a transition from a closed to an open state, and
therefore influence protein function \emph{in vivo}. Despite their
obvious importance, crowding phenomena in membranes are much less well
studied than in the cytoplasm.

Using statistical mechanics results for hard disk liquids, we show
that crowding induces an entropic tension in the membrane, which
influences transitions that alter the projected area and circumference
of a membrane protein.  As a specific case study in this effect, we
consider the impact of crowding on the gating properties of bacterial
mechanosensitive membrane channels, which are thought to confer
osmoprotection when these cells are subjected to osmotic shock. We
find that crowding can alter the gating energies by \ML{more than
  $2\;k_BT$} in physiological conditions, a substantial fraction of
the total gating energies in some cases.

Given the ubiquity of membrane crowding, the nonspecific nature of
excluded volume interactions, and the fact that the function of many
membrane proteins involve significant conformational changes, this
specific case study highlights a general aspect in the function of
membrane proteins.

\pdfbookmark[1]{Author Summary}{Author}
\section*{Author Summary}
Biological membranes are a complex array of lipids and proteins.  The
typical bacterial membrane is made up of hundreds of copies of
different species of membrane proteins embedded in a sea of different
types of lipids.  One of the distinguishing features of biological
matter is the high degree of ``crowding'' to which the different
macromolecules are subjected.  In this work, we explore the
consequences of such crowding in the membrane setting, building upon
earlier work which has primarily focused on how crowding affects
properties in the cytoplasm. The particular case study considered here
centers on a class of membrane channels which respond to tension in
the cell membrane serving to provide osmoprotection to cells subjected
to osmotic shock.  We explore how the critical tension at which these
channels open depends upon the concentration of other membrane
proteins, and conclude that it can be significantly higher at
physiological protein densities compared to the intrinsic value
measured in protein free membranes.

\pdfbookmark[1]{Introduction}{Introduction}
\section*{Introduction}
Cell membranes are packed full of proteins.  The essence of various
membrane inventories is that biological membranes are at least as much
protein as they are lipid. Experiments on the occupancy of biological
membranes by lipids and their protein partners provide a useful basis
for making estimates of the possible consequences of membrane
crowding.  The presence of such high areal fractions of protein means
that there is the possibility that the ``crowding'' effect can alter
the free energies of different membrane protein conformations and the
dynamics of the changes between these conformations as well.  Indeed,
over the last several decades, the importance of crowding effects in
general has become a theme of increasing concern in physical biology
\cite{CR:zimmerman1993,CR:zhou2008,CR:elcock2010,CR:hall2003,CR:rivas2004,CR:zhou2009}.

The question of how the behavior of membrane proteins is altered by
crowding effects has been explored much less thoroughly than their
bulk counterparts\ML{
  \cite{CR:zhou2009,CR:zuckermann2001,CR:aisenbrey2008,CR:leventis2010,CR:minton2000,CR:minton1999}}.
As a concrete example of the way crowding might play out in membranes,
we consider transmembrane proteins that have several conformations
with different areal footprint.  One particularly fascinating class of
proteins of this variety are the mechanosensitive membrane channels.
These proteins are thought to serve as safety valves for cells that
are exposed to osmotic stress, opening up in response to increased
membrane tension for the purpose of equilibrating the cells with their
external environment
\cite{MscL:kung2010,MscL:perozo2006,MscL:perozo2003,MscL:hamill2001}.

To see how crowding might serve as an additional factor in the overall
gating free energy balance for mechanosensitive channels, we consider
the gating tension associated with the mechanosensitive channel of
large conductance (MscL).  Upon opening, at membrane tensions larger
than $\sim 10^{-3}\text{ J/m}^2$, this channel undergoes a change in
radius from roughly \ML{2.4~nm to 3.5~nm}
\ML{\cite{MscL:betanzos2002,MscL:perozo2002,MscL:chiang2004,MscL:martinac2009}}.
As a result of this \ML{increased size, there is a reduction in the
  free area available for the surrounding membrane proteins resulting
  in an entropic driving force to keep the channel closed.}  The work
presented here explores the relative importance of this effect
compared to other contributions to the overall free energy budget for
mechanosensitive channel gating.

In the remainder of this paper, we first examine various estimates of
the degree of crowding in biological membranes. We then go on to
explore the consequences of such gating for the free energy of the
crowded proteins within the membrane, and the accompanying changes of
the channel's gating tension.

\pdfbookmark[1]{Results}{Results}
\section*{Results}
\pdfbookmark[2]{The degree of crowding in membranes}{Degree}
\subsection*{The degree of crowding in membranes}
As a prerequisite to characterizing the functional consequences of
membrane crowding, we must first estimate the extent of crowding found
in different types of membranes.  There are various ways to arrive at
numerical estimates of the extent of crowding of membrane proteins in
biological membranes.  One key measurable quantity that reflects the
fraction of membrane area occupied by proteins is the protein to lipid
mass ratio which typically falls in the range 1-2.5
\cite{schaechter2006,takamori2006,dupuy2008,mitra2004}. Assuming that
transmembrane (TM) domains make up about half of the membrane protein
mass \cite{hahne2008} and have roughly the same density as the lipids
results in the estimate that 30-55\% of the membrane area in the
bilayer plane is occupied by proteins.  Sowers and Hackenbrock
\cite{CR:sowers1981} obtained electron microscopy images of
mitochondrial inner membranes after application of a strong electric
field that made all proteins drift to one end of the membrane surface,
and found that the packed proteins in those images occupy 40-50\% of
the total area.  Ryan \emph{et al.} \cite{CR:ryan1988} fitted a
statistical mechanics model of steric exclusion to the distribution of
fluorescently labeled membrane proteins on rat basophilic leukemia
cells subject to an electric field, and extracted an area coverage of
55-75\%.  Direct experimental estimates of the protein area fraction
in red blood cell plasma membrane and synaptic vesicles have yielded
area fractions of 20-25\% \cite{dupuy2008,takamori2006}.  In an
extreme case, atomic force microscopy images of the photosynthetic
membranes of \emph{Rhodospirillum photometricum} cells
\cite{CR:scheuring05} under various growth conditions show almost
close-packed photosynthetic proteins arranged with nearly crystalline
order.  All of these examples tell the same fundamental story:
membrane proteins are in very close proximity.

Another way of characterizing this crowding is by appealing to the
number density which gives the number of membrane proteins per unit
area of membrane.  Aldea {\it et al.}~\cite{CR:aldea1980} report that
the five major outer membrane proteins (by mass) in \emph{Salmonella
  typhimurium} have a total surface density of about $0.1/\nm^2$ in a
wide range of growth conditions.  Neidhardt {\it et al.}
\cite{neidhardt1990} (p.~41) quote lipoproteins as the most abundant
protein (by number) in \emph{Escherichia coli}, with $\sim 7\times
10^5$ copies in the outer membrane of a typical cell. Estimating the
area of a typical \emph{E.~coli} to be $5\;\mum^2$ \cite{PBoC}, this
gives a density of about $10^5/\mum^2=0.14/\nm^2$.  Another way to
estimate a protein density is to consider the fraction of the genome
that codes for membrane proteins.  In \emph{E.~coli}, about 1/3 of the
4200 genes encode membrane proteins, and the total number of proteins
is about $3\times 10^6$ per cell \cite{PBoC}. If 1/3 of all proteins
are evenly distributed in the two membranes, each membrane has about
500,000 proteins, or about $0.1\text{ protein/nm}^2$.  The areal and
number densities estimated above are roughly consistent. If one
assumes a footprint of $1.5\;\nm^2$ per transmembrane helix
\cite{takamori2006,dupuy2008,mitra2004}, and 3 \ML{transmembrane
  helices} per protein (see below), a number density of $0.1/\nm^2$
corresponds to an area fraction of 0.45.

There are other ways to think about the extent of membrane crowding,
each with its own assumptions and merits, but regardless of these
details the message will be the same. Biological membranes are
crowded!  For the purposes of this article, what these numbers tell us
is that the mean spacing between proteins (estimated by evaluating
$1/\sqrt{c_A}$) is only slightly larger than the proteins themselves,
so that a significant fraction of the membrane area is occupied by
proteins.

\begin{figure}[!t]
\begin{center}
  \includegraphics{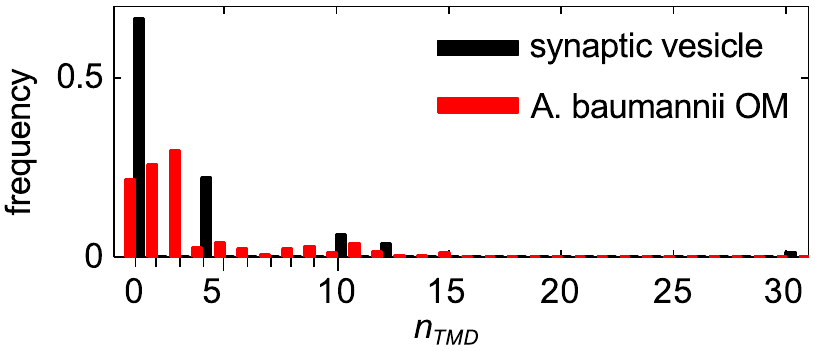} 
\end{center}
  \caption{{\bf Relative abundance of membrane protein subunits with
      different number of transmembrane (TM) helices.} The histograms
    are based on data for synaptic vesicles \cite{takamori2006}, and
    the outer membrane (OM) of the Gram-negative bacterium
    \emph{A. baumannii} \cite{yun2010}. Proteins with no predicted TM
    domains were excluded.}\label{TMHdistribution}
\end{figure}

Membrane proteins are not only abundant, they are also very
heterogeneous, and vary significantly in size and shape
\cite{CR:engelman2005}. Quantitative data on this heterogeneity is
harder to come by\ML{, and we will therefore use the number of
  transmembrane helices $(n_\text{TMH})$ as an approximate
  indicator. Bioinformatic} predictions of transmembrane
regions~\cite{maeller2001} are routinely reported in surveys of
proteins or putative protein-coding DNA
regions~\cite{yun2010,takamori2006,hahne2008}, and range from one to
several tens per protein subunit.  Figure \ref{TMHdistribution} illustrates
two transmembrane helix distributions, based on a synaptic vesicle
model \cite{takamori2006} and a proteomics study of the outer membrane
of the Gram-negative bacterium \emph{Acinetobacter baumannii}
\cite{yun2010}, respectively. The latter is an average of three
different techniques to estimate relative abundance, which differ
significantly in specific cases, but lead to similar overall
distributions (not shown). It is interesting to note the similarities
in distributions in figure \ref{TMHdistribution}, both being dominated by
proteins with a few TM helices, and spanning about one order of
magnitude. However, there are several significant sources of
uncertainty. For example, not all membrane proteins were
detected\cite{yun2010,takamori2006}, and we have not accounted for
aggregation of protein subunits into larger complexes.

In our calculations below, we will \ML{model membrane proteins by
  circular disks}, and will need to estimate
$\eta^2=\Var[R]/\mean{R^2}$, where $\Var[R]=\mean{(R-\mean{R})^2}$
denotes the variance of $R$.  \ML{This quantity, which measures the
  variability of the projected protein area}, enters into the more
sophisticated treatments of the crowding effect discussed later in the
paper.  \ML{A useful approximation is}
$\eta^2\approx\eta^2_\text{TMH}=\Var[\sqrt{n_\text{TMH}}]/\mean{n_\text{TMH}}$,
which \ML{comes from setting $R^2$ proportional to $n_\text{TMH}$.}
Excluding proteins with no predicted transmembrane domains, the
synaptic vesicle and \emph{A. baumannii} outer membrane protein data
sets in figure \ref{TMHdistribution} give 0.25 and 0.14 for
$\eta^2_\text{TMH}$, and 3.0 and 3.5 for the mean number of
\ML{transmembrane helices}, respectively.  As we will see, these
numbers indicate that size variability does not make a large
quantitative contribution to the crowding effect, despite the quite
broad distributions shown in figure \ref{TMHdistribution}.

\pdfbookmark[2]{Crowding effects on gating}{Crowding}
\subsection*{Crowding effects on gating}
In light of estimated membrane protein crowding, our aim is to explore
the implications of such crowding for channel gating.  The total free
energy change upon gating, $\Delta
G_\text{tot}=G_\text{open}-G_\text{closed}$, can be thought of as
arising from multiple contributions.  In particular, we have
\begin{equation}\label{Gterms}
\Delta G_\text{tot}=\Delta G_\text{protein}+\Delta G_\text{load}
+\Delta G_\text{mem}+\Delta G_\text{crowd},
\end{equation}
where the first term reflects the free energy change associated with
the protein degrees of freedom and their internal structural
rearrangements, the second term refers to the potential energy of the
loading device, and the third term characterizes the free energy of
\ML{protein-lipid interactions, including the deformed membrane
  surrounding the protein that} has been implicated as a key player in
the gating of mechanosensitive channels
\cite{phillips2009,wiggins2005,turner2004}. The last term is the
crowding-induced term. A membrane protein with a large cytosolic
domain can potentially be crowded both by molecules in the cytoplasm,
and by other membrane proteins. \ML{While the former effect has in
  fact been observed in the mechanosensitive channel
  MscS\cite{MscS:grajkowski2005}, it is the latter effect that} forms
the main substance of this paper.

\begin{figure}[!t]
\begin{center}
  \includegraphics{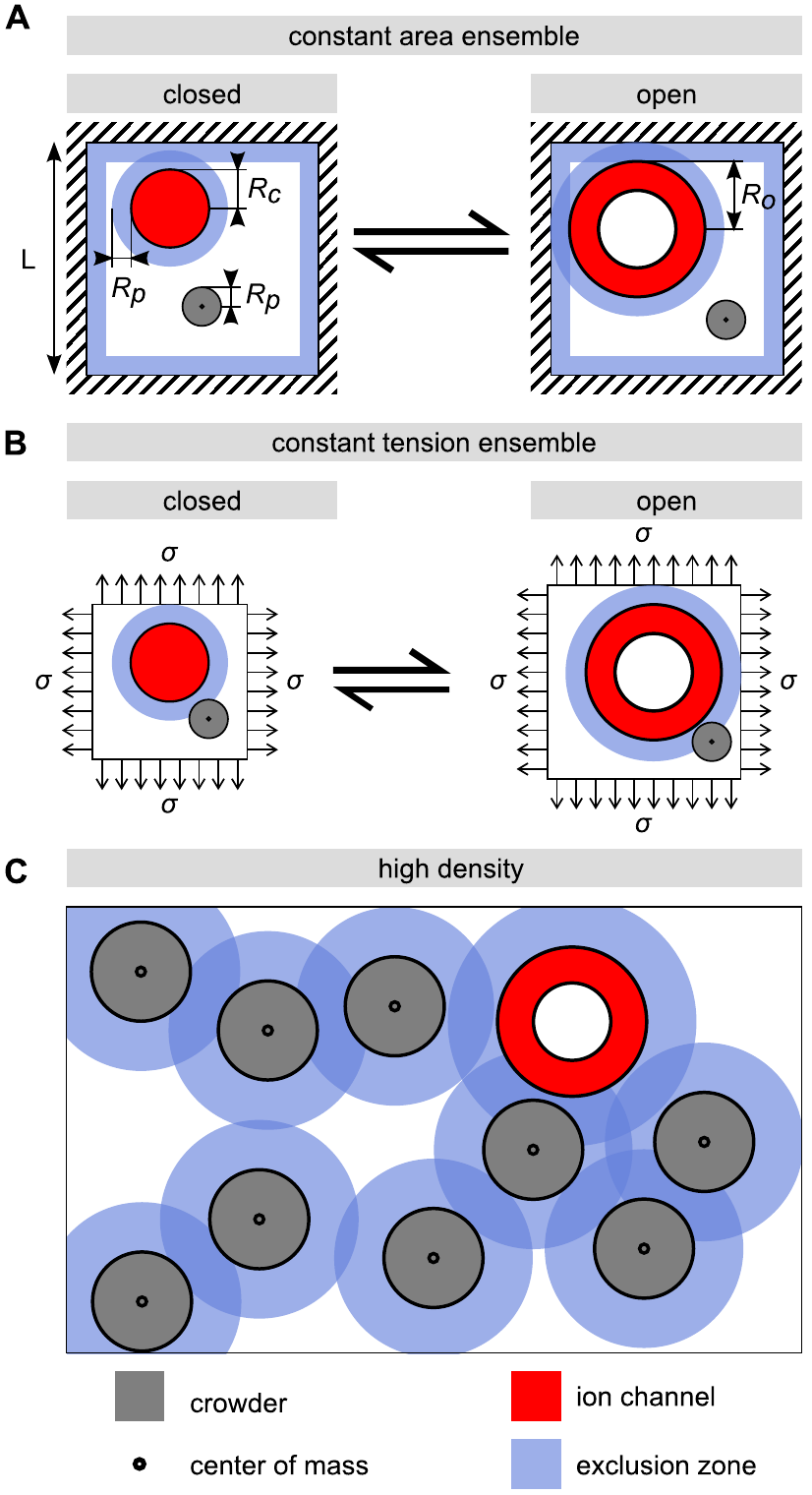}
\end{center}
  \caption{{\bf Excluded-area interactions and channel gating.}
    \ML{(A) Gating of a channel (red) crowded by a single crowder
      (gray) of radius $R_p$ in the constant area ensemble, where the
      total surface area is fixed by the outer walls (dashed). (B) In
      the constant tension ensemble with applied tension $\sigma$, the
      total area increases as the channel opens, so that the total
      lipid area is conserved.} For disk-shaped particles of finite
    size, the free area available for each center of mass is limited
    by the minimum distance between two centers of mass. This effect
    can be illustrated by exclusion zones of width $R_p$ around each
    protein.  \ML{In the constant tension ensemble, the reduced area
      for the crowders is due to larger exclusion zone in the open
      compare to the closed state. In the high density regime (C), the
      exclusion zones overlap, which complicates the analysis. We use
      scaled-particle theory to analyze this case.}}
    \label{ExclusionEntropy}
\end{figure}

The main conceptual point of the remainder of the paper can be stated
simply as the idea that when the channel opens and changes its radius
from ``small'' to ``large'', there will be a free energy cost for the
surrounding membrane proteins which we will refer to as crowders.  In
particular, these crowders will have their entropy reduced, which
amounts to an effective pressure on the channel walls opposing its
opening.  To explore this claim, we will work in two distinct
ensembles.

In the (mathematically) simpler case, we imagine a two-dimensional
membrane ``box'' like that shown in figure
\ref{ExclusionEntropy}\ML{A}, such that the overall area is fixed.
When the channel goes from the closed to the open state, there is a
net reduction in the available area for the remaining crowders, which
results in an entropic tension that favors the closed state.  We make
no reference to the elastic cost of squishing the lipids to access
this state, since it can be shown that this energy is negligible in
comparison with our main contribution of interest which is the
entropic effect (\ML{see supporting text S1, Sec.~1}).

The second scenario imagines a loading device that subjects the
membrane to some fixed tension on its perimeter, much like the springs
that hold a trampoline under its state of tension.  It can be shown
that in this case, when the channel goes from the closed to the open
state, the areal strain\ML{, and hence the lipid area available to the
  crowders, do} not change significantly (\ML{see supporting text S1,
  Sec.~2}).  However, because of the change in the circumference of
the protein, the exclusion annulus around the channel, indicated in
figure \ref{ExclusionEntropy}, will be enlarged. Hence, there will
still be an entropic tension which favors the closed state.  In both
cases, we make the implicit assumption that the number of lipids in
the membrane does not change on the time scale of protein
conformational changes.

To explore these two scenarios, we begin with the box of fixed area
and use the simplest ``ideal gas'' physics to evaluate the change in
entropy due to the loss of translational degrees of freedom when the
channel goes from the closed to the open state.  In particular, the
translational entropy of one crowder can be computed as the logarithm
of the area available to its center of mass,
\begin{equation}
 g_\text{crowd}(R)= -\kBT~\ln 
 \frac{\big(L^2 - \pi(R+R_p)^2-A_\text{edge}\big)
 }{A_\text{lattice}},
 \end{equation}
where $R_p$ is the radius of the crowder\ML{, $R$ is the radius of the
  channel}, $A_\text{edge}$ is the band of thickness $R_p$ around the
edge of the box \ML{from which the crowder center of mass is excluded}
(see figure \ref{ExclusionEntropy}\ML{A}). The denominator
$A_\text{lattice}$ refers to a discretization length scale used in a
lattice model for the entropy \cite{PBoC}. Hence, the numerator is the
effective area available to the crowder in the $L \times L$ membrane
patch, recognizing that the minimal center-of-mass distance between
the crowder and channel is $R_p+R$ (see figure
\ref{ExclusionEntropy}).

This expression can be simplified by expanding in the small parameter
$(\pi(R+R_p)^2+A_\text{edge})/L^2$.  If we exploit this simplification
and add the contributions from $N$ crowders, the difference in free
energy between the open $(R=R_o)$ and closed $(R=R_c)$ states due to
the crowding contributions can be written as $ \Delta G_\text{crowd}
=N\Delta g_\text{crowd}\approx c_A k_BT\big(\pi (R_o^2 -R_c^2)+2\pi
R_p (R_o-R_c)\big)$, with $c_A=N/L^2$ being the crowder
concentration.  These two terms have simple and intuitive
interpretations that are serviced by noting that we can rewrite the
area and circumference change, respectively, as $\Delta A=\pi (R_o^2
-R_c^2)$ and $\Delta C=2\pi (R_o-R_c)$. We can then divide the
entropic crowding tension into a surface and a line tension, and write
  \begin{equation}\label{crowdingtensions}
    \Delta G_\text{crowd}\approx 
    -\sigma_\text{crowd}\Delta A
    +\tau_\text{crowd}\Delta C,
  \end{equation}
in the \ML{constant} area ensemble.  In our ``ideal gas''
approximation, the surface tension is $-\sigma_\text{crowd}=c_A\kBT$,
the familiar ideal gas law. The line tension,
$\tau_\text{crowd}=c_AR_p\kBT$, originates in the fact that the
annulus of exclusion shown in figure \ref{ExclusionEntropy} changes
size upon gating. This contribution vanishes in the limit that the
size of the crowders goes to zero.  For both terms, we will need to
appeal to our earlier estimates of protein areal concentrations to set
the scale of the effect.

We can now consider the second scenario in which there is a fixed
applied tension $\sigma$\ML{, shown in figure
  \ref{ExclusionEntropy}B}.  \ML{Neglecting edge effects, the lipid
  area in which the crowders wiggle around does not change in this
  case}, but the annulus of exclusion does, and hence the contribution
of the entropy change to the free energy is given by the $\Delta C$
term only, i.e.,
  \begin{equation}\label{crowdingtensionsCT}
    \Delta G_\text{crowd}\approx 
    \tau_\text{crowd}\Delta C.
  \end{equation} 
At the same time, there is a relaxation in the energy of the loading
device which takes the form $\Delta G_\text{load}=-\sigma\Delta A$ in
the \ML{constant} tension ensemble.

The treatment given above provides the simplest estimate of the
crowding effect.  However, as shown in figure
\ref{ExclusionEntropy}\ML{C}, things become more complicated in the
high concentration limit.  In particular, the amount of available area
is much less than is suggested by the simple estimate above, where we
made no reference to the way the crowders interact with each
other. Neglecting these interactions underestimates the crowding
effects. For 50\% protein area coverage, the more accurate
computations described below give increased surface and line tension
terms by a factor of four and two, respectively.  Note that the
entropic effect increases in a highly non-linear fashion with the
crowder area fraction, effectively diverging as one reaches the
closed-packing limit. The effect we describe can thus be potentially
much larger than the already substantial estimates of $1-10\,\kBT$
summarized below.

One way to think about this, illustrated in figure
\ref{ExclusionEntropy}\ML{C}, is in terms of exclusion zones around
each crowder, analogous to the physics described by the van der Waals
theory of gases.  In the highly crowded regime, the theoretical
difficulty is to compute the total size of the exclusion zones in a
way that avoids double counting areas where multiple exclusion zones
overlap. We use scaled-particle theory for mixtures of hard disks, an
approximate equation of state that combines reasonable accuracy with
analytical tractability
\cite{SPT:reiss59,SPT:helfand61,SPT:lebowitz65}, and has been widely
applied to describe the effects of crowding
\cite{CR:zhou2009,CR:zimmerman1993,CR:zhou2008,CR:zuckermann2001,CR:aisenbrey2008,CR:chatelier1996,CR:minton1999}.
The central results, for circular crowders, are presented in table
\ref{thetable}, in terms of the concentration $c_A$, areal fraction
$\phi$, and, for non-uniform crowder size, relative size variance
$\eta^2=(\mean{R_p^2}-\mean{R_p}^2)/\mean{R_p^2}$. Details of the
derivations are presented in the Models section below.  The
crowding-induced changes in gating energy still take the form of
Eqs.~\eqref{crowdingtensions} and \eqref{crowdingtensionsCT}, with
only the line tension contributing in the constant tension ensemble.
The more exact scaled particle theory gives larger crowding tensions.
\begin{table}[!t]
\caption{\bf{Entropic surface and line tensions induced by crowders}}\label{thetable}
\begin{center}
    \begin{tabular}{|l|c|c|}
      \hline
      & $-\sigma_{\mathrm{crowd}}/\kBT$&$\tau_{\text{crowd}}/\kBT$\\
      \hline
      ideal gas& $c_A$ & $c_A R_p$\\
      \hline
      SPT, uniform crowders&
      $\frac{c_A}{(1-\phi)^2}$&$\frac{c_AR_p}{1-\phi}$\\
      \hline
      SPT, non-uniform crowders&
      $\frac{c_A(1-\phi\eta^2)}{(1-\phi)^2}$&$\frac{c_A\mean{R_p}}{1-\phi}$\\
      \hline
    \end{tabular}
\end{center}
\begin{flushleft}Entropic surface and line tensions induced by crowders,
    estimated by an ideal gas calculation and scaled-particle theory
    (SPT, see Eqs.~\eqref{dFconstantarea} and \eqref{dGresult}).  The
    results are derived for the case in which a single circular
    protein increases its radius from $R_c$ to $R_o$ in the presence
    of circular crowders with radius $R_p$. The non-uniform crowders
    case contains averages $\mean{\cdot}$ over the crowder radius
    distribution, and this size variation
    $(\eta^2=\Var[R_p]/\mean{R_p^2}\ge 0)$ leads to a smaller surface
    tension effect compared to uniform crowders with the same mean
    size.
\end{flushleft}
 \end{table}

With these analytical results in hand, we now turn to the question of
the actual magnitude of the crowding effect.  To be concrete, we
consider the case in which we have a membrane where the area is half
lipids and half proteins (i.e. $\phi = 1/2$), big enough to make the
ideal gas estimates questionable.  We consider a radius change of a
single channel from \ML{2.4 to 3.5 nm (as is appropriate for MscL
  \cite{MscL:chiang2004})}, and a crowder radius of 1 nm.  For
simplicity, we also neglect size variability and set $\eta^2=0$ since
using our estimate of $\eta^2_\text{TMH}$ in the range 0.15-0.25 would
only give a $\sim 10\;\%$ correction. This leads to an estimated
crowder density of $c_A=\phi/\pi R_p^2\approx 0.16\;\text{nm}^{-2}$.
Using these numbers in the context of table \ref{thetable}, we get
$-\sigma_\text{crowd,SPT}\approx 0.64\;\kBT/\text{nm}^2$, and
$\tau_\text{crowd,SPT}\approx 0.32\;\kBT/\text{nm}$. \ML{This}
translates to crowding-induced changes in total gating energies of
\ML{15.2} and \ML{2.2} $\kBT$, for the constant area and constant tension
ensembles, respectively (see table \ref{numbertable}). \ML{Even in the
  case of constant tension, most relevant to MscL, the crowding effect
  can have a sizable impact on channel gating.}
\begin{table}[!t]
\caption{\bf{Estimated crowding effects on MscL gating.}}\label{numbertable}
\begin{center}
\begin{tabular}{|l|c|c|c|c|c|}
\hline
&\multicolumn{2}{c|}{\ML{constant} area}& 
\multicolumn{2}{c|}{\ML{constant} tension}&\\
\hline
&IG&SPT&IG&SPT&units\\
\hline
$\Delta G_{\text{crowd}}$& 4.3 & 15.2 & 1.1 & 2.2 &$\kBT$\\
\hline
$\Delta\sigma_\text{crowd}$&0.21& 0.74 & 0.05& 0.11&$\frac{\kBT}{\nm^2}$\\
\hline
\end{tabular}
\end{center}
\begin{flushleft}
Different metrics for the effect of crowding on the gating behavior of
a mechanosensitive channel.  The first row shows the approximate
\ML{changes in} gating energies. The second row shows the
corresponding increase in gating tension,
$\Delta\sigma_\text{crowd}=\Delta G_\text{crowd}/\Delta A$, which can
be measured directly in patch-clamp experiments. For comparison, the
typical gating tension for isolated MscL is 0.3-1.3 $\kBT$/\nm$^2$.
\ML{For MscL gating, the constant tension ensemble is the more
  appropriate model.}
\end{flushleft}

\end{table}

\begin{table}[t!]
\caption{\bf{List of symbols.}}\label{notation}
\begin{tabular}{|c|l|l|}
  \hline
  $n_\text{TMH}$ & number of transmembrane helices&\\
  \hline
  $\beta$ & inverse temperature scale &$\beta=1/\kBT$\\
  \hline
  $\vec{n}$ & copy number vector& $\vec{n}=(n_1,n_2,\ldots)$\\
  \hline
  $n$& total copy number &$n=\sum_j n_j$\\
  \hline
  $\vec{N}$ & copy number vector for the crowders& $\vec{N}=(N_1,N_2,\ldots)$\\
  \hline
  $N$& total number of crowders&$N=\sum_{j\ne o,c}N_j$\\
  \hline
  $A$& area&\\
  \hline
  $\sigma$& surface tension (2D analog of negative pressure)&\\
  \hline
  $c_A$& areal number density &$c_A=n/A$\\
  \hline
  $R_i$& \parbox[t]{0.55\textwidth}{In-plane radius of species $i$. In particular, $i=o,c$ 
  for the open and closed channel conformations respectively.}&\\
  \hline
  $F$ & Helmholtz free energy (constant area ensemble)&\\
  \hline
  $G$ & Gibbs free energy (constant tension ensemble)& $G=F-\sigma A$\\
  \hline
  $\Delta C$& circumference change  of channel& $\Delta C=2\pi(R_o-R_c)$\\
  \hline
  $\Delta A$& area change  of channel& $\Delta A=\pi(R_o^2-R_c^2)$\\
  \hline
  $\mean{R^m}$ & $m:th$ moment of the protein radius distribution& $\mean{R^m}=\frac{1}{n}\sum_jn_j R_j^m$\\
  \hline
  $\Var[R]$& radius variance& $\Var[R]=\mean{R^2}-\mean{R}^2$\\
  \hline
  $\mean{R_p^m}$ &   $m:th$ moment of the crowder radius distribution & $\mean{R_p^m}=\frac{1}{N}\sum_{j\ne o,c}N_j R_j^m$\\
  \hline
  $R_p$ & crowder radius, for the case of uniform crowder size&\\
  \hline
  $\phi$& area fraction of disks or proteins& $\phi=c_A\pi\mean{R^2}$\\
  \hline
  $A_0$& total area occupied by unstretched lipids &$A_0=A(1-\phi)=A-n\pi\mean{R^2}$\\
\hline
  $\eta^2$& relative protein radius variance &$\eta^2=\Var[R]/\mean{R^2}$\\
\hline
  $A_{\vec{n}}$ & unstretched equilibrium area in constant tension ensemble&
  $A_{\vec{n}}=A_0+n\pi\mean{R^2}$.\\  
\hline
\end{tabular}
\end{table}

\pdfbookmark[1]{Models}{Models}
\section*{Models}
In this section, we review some basic results of scaled-particle
theory, and derive the main results of \ML{the previous section and}
table \ref{thetable}.  Motivations for some of the approximations we
use, such as neglecting the area compression of the lipid bilayer, are
given in the supporting text S1. See table \ref{notation} for a
summary of the notation and symbols.

\pdfbookmark[2]{Basic results of scaled particle theory}{Basic}
\subsection*{Basic results of scaled particle theory}\label{basicSPT}
In this section, we restate some basic results of scaled-particle
theory that serve as the basis for \ML{our calculations}.  We will
express the results in terms of area, temperature, and particle copy
numbers, since this is what we will use as thermodynamic control
variables, but also quote the results in terms of variables like
concentration $c_A$ and area fraction $\phi$ (see table
\ref{notation}).

Scaled-particle theory has been generalized to heterogeneous mixtures
of convex particles \cite{SPT:gibbons69,SPT:boublik75,SPT:talbot1997},
but we restrict our attention to mixtures of circular disks
\cite{SPT:lebowitz65}. We will simply quote the results we need, and
refer to the literature for details on the derivations
\cite{CR:zhou2009,SPT:lebowitz65,SPT:helfand61,SPT:reiss59,SPT:heying04}.

We start with the canonical partition function for a collection of
hard disks with radii $R_j$ and copy numbers $n_j$, enclosed in an
area $A$. The crowding effect we are interested in comes from the
configurational entropy of the proteins, and we therefore omit
velocities and internal degrees of freedom, and neglect boundary
effects.  The remaining configurational partition function depends on
the many-particle interaction energy, $U(\{\vec{x}\})$,
\begin{equation}\label{Zdef0}
  Z(\vec{n},A,T)=\frac{1}{\prod_jn_j!}
  \int d^n\vec{x}\;e^{-\beta U(\{\vec{x}\})},
\end{equation}
where $\beta=1/\kBT$ is the inverse temperature, $j$ is the disk
species index, and we use vector notation $\vec{n}=(n_1,n_2,\ldots)$
to denote the copy number distribution, with $n=\sum_j n_j$ being the
total number of disks (see also table \ref{notation}).  We next factor
$Z$ by multiplying and dividing by $A^n$, and write
\begin{equation}\label{Zdef}
  Z(\vec{n},A,T)=Q(\vec{n},A,T)\prod_j\frac{A^{n_j}}{n_j!},
\end{equation}
where $Q=\frac{1}{A^n}\int d^n\vec{x}\;e^{-\beta U(\{\vec{x}\})}$
describes the deviation from ideal gas behavior due to the interaction
energy $U$, which we take to be simple hard-disk repulsion. (By
construction, $Q=1$ for an ideal gas, where $U=0$ for all
configurations.)

For the computations below, we will break down configurational changes
as removals and insertions of particles of different sizes, and
also consider area changes as a result of changed particle size. We will
therefore need the chemical potential and surface tension of the disk
mixture that is our protein model.

Scaled particle theory offers a simple equation of state that relates
the surface tension (2D analog of negative pressure) exerted by the
disks to the area footprint, number density, and size variation of the
disks.  Rewriting for example Eq.\ (6.7) of
ref.\ \cite{SPT:lebowitz65} in our notation, we get
\begin{equation}\label{SPTtension0}
  \frac{\sigma_\text{SPT}}{\kBT} =-\fixdiff{\ln Z}{A}_{T,\vec{n}}
  =-\frac{c_A}{1-c_A\pi\mean{R^2}}
  -\frac{\pi(c_A\mean{R})^2}{(1-c_A\pi\mean{R^2})^2}.
\end{equation}
(Note that we use the sign convention $\sigma dA$ for surface
tension-area work, which is the opposite sign compared to the
pressure-volume convention $-pdV$ used in the original derivations of
scaled particle theory).  After substituting $c_A=n/A$, this
expression can be brought to the more compact form
\begin{equation}\label{SPTtension1}
  \frac{\sigma_\text{SPT}}{\kBT} 
 =-n\frac{A-n\pi\Var[R]}{\big(A-n\pi\mean{R^2}\big)^2},
\end{equation}
where $\Var[R]=\mean{R^2}-\mean{R}^2$ is the disk radius
variance. Note how the size variability decreases the (negative)
pressure through the variance term.

This can again be rewritten in terms of concentration, area fraction,
and relative size variability by combining the concentration $c_A=n/A$
and the area fraction relation $A-n\pi\mean{R^2}=(1-\phi)A$, and then
the relation
$\frac{n\pi\Var[R]}{A}=c_A\pi\mean{R^2}\eta^2=\phi\eta^2$, resulting
in
\begin{equation}\label{SPTtension2}
 \frac{\sigma_\text{SPT}}{\kBT} 
 =-\frac{c_A}{(1-\phi)^2}\Big(1-\frac{n\pi\Var[R]}{A}\Big)
 =-\frac{c_A(1-\phi\eta^2)}{(1-\phi)^2}.
\end{equation}
When we consider area changes in the next section, the area integral
of the surface tension at constant particle numbers will also come in
handy, and we therefore integrate \Eq{SPTtension1}, and obtain
\begin{equation}\label{SPTtensionIntegral}
  \int\frac{\sigma_\text{SPT}}{\kBT}dA =
  -n\ln\big(A-n\pi\mean{R^2}\big)
  +\frac{n^2\pi\mean{R}^2}{A-n\pi\mean{R^2}}.
\end{equation}

Finally, we will need the chemical potential. This is commonly divided
into an ideal gas part plus a correction, called the excess chemical
potential. Using $\vec{n}+\hat{e}_j$ to denote the state with an added
particle of species $j$, the excess chemical potential is defined as
the ratio
\begin{equation}\label{deltaMu1}
  \frac{\Delta\mu_j}{\kBT}
   =-\ln\frac{Q(\vec{n}+\hat{e}_j,T,A)}{Q(\vec{n},T,A)}.
\end{equation}
Using manipulations similar to those that lead to
Eqs.~\eqref{SPTtension0} and \eqref{SPTtension1}, the scaled-particle
theory approximation given by, e.g.,
refs.\ \cite{CR:zhou2009,SPT:lebowitz65}, can be rewritten in the form
 \begin{equation}\label{deltaMu2}    \frac{\Delta\mu_j}{\kBT}
   =-\ln\Big(1-\frac{n\pi\mean{R^2}}{A}\Big)
   +\frac{n(\pi R_j^2+2\pi R_j\mean{R})}{A-n\pi\mean{R^2}}
   +\left(\frac{n\pi R_j\mean{R}}{A-n\pi\mean{R^2}}\right)^2,
 \end{equation}
where the averages should be computed with copy numbers $\vec{n}$,
i.e., without the test particle present.

From the definition of $\Delta\mu$ (\Eq{deltaMu1}), one can see that
$e^{-\beta\Delta\mu}=\frac{Q(\vec{n}+\hat{e}_j,T,A)}{Q(\vec{n},T,A)}$
also has a probabilistic interpretation, namely as the probability
that a test particle can be inserted somewhere in the fluid without
overlapping with the other particles. This observation, which is
exact, is in fact the starting point for one way to derive
scale-particle theory (see e.g., \cite{CR:zhou2009,SPT:lebowitz65}),
by using a clever approximation to account for the overlapping
exclusion zones in figure \ref{ExclusionEntropy}\ML{C}.

Finally, the chemical potential is given by the ratio of
partition functions,
\begin{equation}
  \frac{\mu_j}{\kBT}
  =-\ln\frac{Z(\vec{n}+\hat{e}_j,T,A)}{Z(\vec{n},T,A)}.
\end{equation}
If we substitute \Eq{Zdef} and then \Eq{deltaMu2}, we get the
scaled-particle approximation to the chemical potential, namely,
\begin{equation}\label{fullMu}
  \frac{\mu_j}{\kBT}
  =-\ln\frac{A^{n_j+1}}{(n_j+1)!}\frac{n_j!}{A^{n_j}}+\Delta\mu_j
  =-\ln\Big(\frac{A-n\pi\mean{R^2}}{n_j+1}\Big)
  +\frac{n(\pi R_j^2+2\pi R_j\mean{R})}{A-n\pi\mean{R^2}}
  +\left(\frac{n\pi R_j\mean{R}}{A-n\pi\mean{R^2}}\right)^2.
\end{equation}

\pdfbookmark[2]{Gating transition in the constant area ensemble}{Gating}
\subsection*{Gating transition in the constant area ensemble}
With the results of the previous section in hand, we are now in a
position to derive our main results.  Specifically, we will consider a
situation with a single channel crowded by other proteins that do not
change their configuration. We denote the copy number vector and total
number of these crowders by $\vec{N}$ and $N$ respectively. The state
with a channel in state $i$ ($i=o,c$ for the open and closed state
respectively) will then have the copy number vector
$\vec{N}+\hat{e}_i$.

In the results and discussion sections, we use $G$ to denote a generic
free energy. In the following derivations, we will be more precise,
and use $F$ and $G$ for the free energy in the constant area and
constant tension ensembles, respectively. In the thermodynamic limit,
they are related by a Legendre transformation
$G(\vec{n},\sigma,T)=F(\vec{n},A,T)-\sigma A$, where $\sigma$ is the
surface tension.

When computing the gating energy changes, we expand in various small
parameters. Specifically, we will consider the total area, or total
lipid area, to be much larger than the area of a single protein of any
species, but comparable to the total crowder footprint
$N\pi\mean{R^2}$. This means that $\pi R_j^2/A$ is a small parameter
for all protein radii $R_j$, but $N\pi R_j^2/A \;(=\phi)$ is not
small. In a typical \emph{E.~coli} cell, $A=5$ $\mum^2$, which means
that $\pi R_j^2/A \sim 10^{-6}$ (for $R_o=3.5$ nm).  We will neglect
such small terms.

To compute the free energy changes of a conformational change at
constant total area, e.g., changing a particle from species $i$ (say,
a closed channel) to species $j$ (say, an open channel), we subdivide
the reaction into one insertion and one removal, by multiplying and
dividing by the partition function of the intermediate state,
\begin{equation}
  \frac{\Delta F_{i\to j}}{\kBT}=
  -\ln\frac{Z(\vec{N}+\hat{e}_j,A,T)}{Z(\vec{N}+\hat{e}_i,A,T)}
  =-\ln\Big(\frac{Z(\vec{N}+\hat{e}_j,A,T)}{Z(\vec{N},A,T)}
  \frac{Z(\vec{N},A,T)}{Z(\vec{N}+\hat{e}_i,A,T)}\Big).
  \end{equation}
Splitting the product of ratios, we can compare with \Eq{fullMu} to
identify the free energy change as the difference of chemical
potentials for the two configurations,
\begin{equation}
  \frac{\Delta F_{i\to j}}{\kBT}=
  \underbrace{-\ln\frac{Z(\vec{N}+\hat{e}_j,A,T)}{Z(\vec{N},A,T)}
  }_{\mu_j(\vec{N},A,T)/\kBT}
  \underbrace{+\ln\frac{Z(\vec{N}+\hat{e}_i,A,T)}{Z(\vec{N}+\hat{e}_i,A,T)}
  }_{-\mu_i(\vec{N},A,T)/\kBT}.
\end{equation}
This means that we can use \Eq{fullMu} with $j=o,c$ to compute the
entropic contribution to the free energy change. Using $R_p$ to denote
crowder radii, we get
\begin{align}
  \frac{\Delta F}{\kBT}=&\frac{\mu_o(\vec{N},A,T)-\mu_c(\vec{N},A,T)}{\kBT}\\
  =&\frac{N\pi(R_o^2-R_c^2)+2N\pi\mean{R_p}(R_o-R_c)}{A-N\pi\mean{R_p^2}}
  +\frac{N\pi\mean{R_p}^2\times N\pi(R_o^2-R_c^2)}{(A-N\pi\mean{R_p^2})^2}.
\end{align}
To simplify, we first identify changes in area $\Delta
A=\pi(R_o^2-R_c^2)$, and circumference $\Delta C=2\pi(R_o-R_c)$,
\begin{equation}
  \frac{\Delta F}{\kBT}= \frac{N\mean{R_p}\Delta C}{A-N\pi\mean{R_p^2}}
  +\frac{N\Delta
    A}{A-N\pi\mean{R_p^2}}\bigg(1+\frac{N\pi\mean{R_p}^2}{A-N\pi\mean{R_p^2}}\bigg).
\end{equation}
Next, we use the same simplifications that lead to \Eq{SPTtension2},
and end up with
\begin{equation}\label{dFconstantarea}
  \frac{\Delta F}{\kBT}=
  \frac{c_A\mean{R_p}}{1-\phi}\Delta C
  +\frac{c_A(1-\phi\eta^2)}{(1-\phi)^2}\Delta A.
\end{equation}
The coefficients of $\Delta C$ and $\Delta A$ are the line and surface
tensions tabulated on line three of table \ref{thetable}. The negative
surface tension of the crowders (\Eq{SPTtension2}) acts to oppose an
increased radius of the protein, because increasing the protein
footprint decreases the area available to the rest of crowders.  The
quantities in these coefficients should be computed without the
channel present (although computing them with the channel present
would only make a small difference). The properties of the channel
itself only enter through $\Delta C$ and $\Delta A$.  We obtain the
uniform crowders result (line 2 of table \ref{thetable}) as a special
case, by replacing the mean radius by a single value, $\mean{R_p}\to
R_p$, and set the coefficient of variation, $\eta^2$, to zero.

Next, we consider the constant tension ensemble, and show that we
recover only the line tension effect, i.e., the $\Delta C$ term, in
that case.

\pdfbookmark[2]{Gating transition in the constant tension ensemble}{Gating2}
\subsection*{Gating transition in the constant tension ensemble}
For the constant tension ensemble, the statistical mechanics recipe is
to introduce an external tension $\sigma$, i.e., an external loading
device with energy $-\sigma A$. We also include a term
$H_\text{lipids}$ for lipid elastic energy as a function of area, and
integrate over all areas,
\begin{equation}\label{CTpartition}
  \Xi(\vec{n},\sigma,T)=\int dA e^{\beta(\sigma A -H_\text{lipids})}Z(\vec{n},A,T).
\end{equation}
As we show in the supporting text S1, real membranes are too stiff
for changes and fluctuations in lipid area to give significant
contributions to the gating energy of a single channel. This means
that the above integral will be dominated by the area
$A_{\vec{n}}=A_0+n\pi\mean{R^2}$, where $A_0$ is the total unstretched
lipid area.  To good approximation, we can therefore set $e^{-\beta
  H_\text{lipids}}\approx\delta (A-A_{\vec{n}})$, and think of the
lipids as having constant area and infinite stiffness. This makes it
easy to evaluate the area integral,
\begin{equation}\label{deltaApprox}
  \Xi(\vec{n},\sigma,T)\approx Z(\vec{n},A_{\vec{n}},T)e^{\beta\sigma A_{\vec{n}}},
\end{equation}
and we recover the free energy of the constant tension
  ensemble as the Legendre transformation of the free energy for the
constant area ensemble,
\begin{equation}\label{XiLegendre}
  \underbrace{-\kBT\ln\Xi(\vec{n},\sigma,T)}_{G(\vec{n},\sigma,T)}
  \approx
  \underbrace{-\kBT\ln Z(\vec{n},A_{\vec{n}},T)-\sigma A_{\vec{n}}
  }_{F(\vec{n},A_{\vec{n}},T)-\sigma A_{\vec{n}}}.
\end{equation}

We now return to our test problem, and again denote the crowder copy
numbers by $\vec{N}$, the presence of a channel in state $i=o,c$ by
$\vec{N}+\hat{e}_i$ etc.  We can then divide the total free energy
change into three contributions: removal of a closed channel at area
$A_{\vec{N}+\hat{e}_c}$, an overall area change
$A_{\vec{N}+\hat{e}_c}\to
A_{\vec{N}+\hat{e}_o}=A_{\vec{N}+\hat{e}_c}+\Delta A$ with no channel
present, and insertion of an open channel at area
$A_{\vec{N}+\hat{e}_o}$:
\begin{align}\label{dGcontributions}
  \Delta G
  =&\underbrace{-\mu_c(\vec{N},A_{\vec{N}+\hat{e}_c},T)}_{\text{Removing a closed channel.}}
  \underbrace{-\sigma\Delta A
  +\int_{A_{\vec{N}+\hat{e}_c}}^{A_{\vec{N}+\hat{e}_o}}\sigma_\text{SPT}(\vec{N})dA
  }_{\text{Area change with the channel absent.}}
  \underbrace{+\mu_o(\vec{N},A_{\vec{N}+\hat{e}_o},T)}_{\text{Inserting an open channel.}}.
\end{align}
Substituting \Eq{SPTtensionIntegral} and \Eq{fullMu}, and assuming
that the crowder background does not contain any other channels
$(N_o=N_c=0)$, we get (after collecting terms)
  \begin{multline}\label{dGtension1}
    \frac{\Delta G}{\kBT}=
    -\ln\big(A_0+\pi R_o^2\big)
    +\ln\big(A_0+\pi R_c^2\big)-\frac{\sigma\Delta A}{\kBT} 
    +\frac{N\pi(R_o^2+2\mean{R_p}R_o)}{A_0+\pi R_o^2}
    -\frac{N\pi(R_c^2+2\mean{R_p}R_c)}{A_0+\pi R_c^2}\\
    +\Big(\frac{N\pi\mean{R_p}R_o}{A_0+\pi R_o^2}\Big)^2
    -\Big(\frac{N\pi\mean{R_p}R_c}{A_0+\pi R_c^2}\Big)^2
    -N\ln\Big(\frac{A_0+\pi R_o^2}{A_0+\pi R_c^2}\Big)
    +\frac{N^2\pi^2\mean{R_p}^2(R_o^2-R_c^2)}
    {A_0^2\big(1+\pi R_o^2/A_0\big)\big(1+\pi R_c^2/A_0\big)}.    
  \end{multline}
Next, we Taylor expand in the small parameters $\pi R_{o,c}/A_0$,
collect coefficients of $\Delta C$ and $\Delta A$ (most of which
cancel), and end up with the following lowest order result:
\begin{equation}
  \frac{\Delta G}{\kBT}
  =\frac{N\mean{R_p}\Delta C}{A_0}
  -\frac{\sigma\Delta A}{\kBT} +\text{small terms},
\end{equation}
Noting that $N/A_0=c_A/(1-\phi)$ and discarding the small terms,
we finally get
\begin{equation}\label{dGresult}
  \frac{\Delta G}{\kBT}
  =\frac{c_A\mean{R_p}\Delta C}{1-\phi}-\frac{\sigma\Delta A}{\kBT}.
\end{equation}
Comparing with the constant area result of \Eq{dFconstantarea}, we see
that the contribution from the crowding surface tension has canceled,
but that the coefficient of $\Delta C$ is the same, namely the line
tension in table \ref{thetable}.  The extra term $-\frac{\sigma\Delta
  A}{\kBT}$ reflects the work done by the loading device during the
area change, and is independent of crowding conditions.

\pdfbookmark[1]{Discussion}{Discussion}
\section*{Discussion}
Membrane proteins in cellular membranes are crowded. Estimates based
on data from a broad range of organisms and experimental techniques
\cite{schaechter2006,takamori2006,dupuy2008,mitra2004,hahne2008,CR:sowers1981,CR:ryan1988,CR:scheuring05,CR:aldea1980}
indicate that membrane proteins occupy area fractions ranging from
20\% to well over 50\% in different cell membranes. Crowding induces
an entropic tension in the membrane, which favors membrane protein
conformations with smaller areal footprint and circumference. This
effect can be understood qualitatively using simple free area
arguments, but quantitatively meaningful estimates require more
sophisticated theories. We have used scaled-particle theory for hard
disk mixtures
\cite{SPT:reiss59,SPT:helfand61,SPT:lebowitz65,CR:zhou2009,CR:zuckermann2001,SPT:heying04,CR:aisenbrey2008}
to compute the crowding induced line and surface tensions (see Models,
Eqs.~\eqref{dFconstantarea} and \eqref{dGresult}, and table
\ref{thetable}). As a case study, we apply these results to estimate
the influence of crowding on the gating tension of the bacterial
mechanosensitive channel MscL. This channel is thought to act as a
safety valve for cells under osmotic stress, opening up in response to
high membrane tension in order to avoid membrane rupture
\cite{MscL:kung2010,MscL:perozo2006,MscL:perozo2003,MscL:hamill2001}.

\begin{figure}[!t]
\begin{center}
\includegraphics{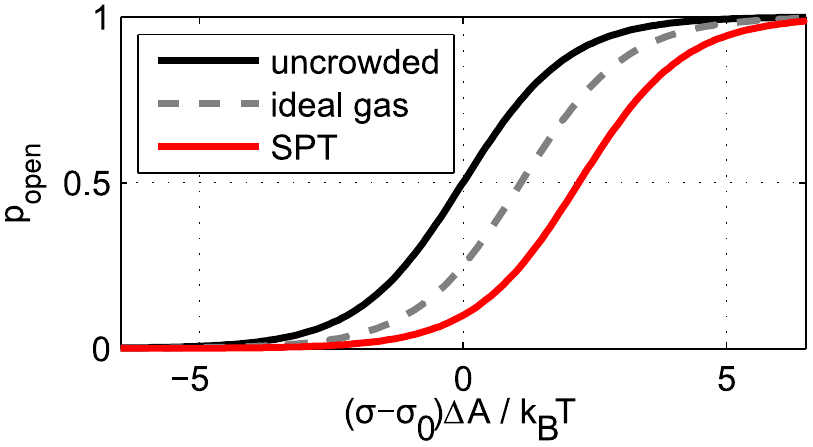}
\end{center}
  \caption{{\bf The effect of crowding on the open probability as a
      function of applied tension $\sigma$.} The graphs illustrate the
    ideal gas (\ML{$\Delta G_\text{crowd}=1.1 \kBT$}) and scaled-particle
    theory (SPT, \ML{$\Delta G_\text{crowd}=2.2\kBT$}) results of table
    \ref{numbertable}, using the constant tension ensemble \ML{as is
      appropriate for MscL}.  All non-crowding contributions to the
    gating free energy are lumped together in the gating tension
    $\sigma_0$.}
      \label{Popen}
\end{figure}

There are different ways to quantify the influence of crowding on
gating behavior (see table \ref{numbertable}).  One way to present the
significance of our results is by appealing directly to the curves
that provide the probability of channel opening as a function of the
driving force.  For the case of a ``two-state'' channel, which
transitions back and forth between distinct closed and open states,
the open probability is $p_\text{open}=(1+\exp(\Delta
G_\text{tot}/\kBT))^{-1}$, where $\Delta G_\text{tot}$ is the energy
difference between the closed and open states, and depends upon the
driving force (such as tension, voltage or ligand concentration).  In
our case, the driving force is the tension, and we can rewrite $
\Delta G_\text{tot} = \Delta G_0-\sigma \Delta A + \Delta
G_\text{crowd}$.  The first term corresponds to all contributions of
\Eq{Gterms} that do not depend explicitly on crowding or applied
tension.  We can rewrite it in the simpler form $\Delta
G_0=\sigma_0\Delta A$.  Figure \ref{Popen} shows the gating
probability $p_\text{open}$ as a function of $\sigma$ both for a
single isolated channel and for the case in which crowders are
present.

An alternative way to decide if the effect is big or small, is to
compare it to some reference energy (or tension). The first relevant
energy scale for comparison is the thermal energy $\kBT$, the energy
scale in the Boltzmann weight $\exp(-\Delta G/\kBT)$ in the open
probability above.  Our numerical examples in table \ref{numbertable}
all change the gating free energy by $\ge 1\;\kBT$.  A second relevant
energy scale is that associated with the gating of various
mechanosensitive channels.  The gating properties of channels such as
MscL have been measured using several different species of lipid
molecules in the surrounding membrane.  The outcome of these elegant
experiments is that the gating energies have typical values of
5-20~$k_BT$ \cite{MscL:perozo2002,MscL:sukharev1999} and corresponding
gating tensions in the range of $0.3-1.3\;k_BT/\nm^2$. In the presence
of spontaneous curvature inducing lipids, these energies and tensions
are even smaller (or even negative, meaning that the channel opens
spontaneously without any applied tension) \cite{MscL:perozo2002}. The
change in gating tension due to crowding is $\Delta
\sigma_\text{crowd}= \Delta G_\text{crowd}/\Delta A$, and we get
numbers in the range $0.05-0.7\;k_BT/\nm^2$.

The entropic cost of channel opening in a crowded solution of membrane
proteins has so far been discussed only with reference to hard core
repulsion between proteins. It is however well known that
membrane-mediated interactions may emerge from the overlap of the
membrane deformations surrounding neighboring proteins, such as those
arising from a thickness mismatch between the hydrophobic protein core
and the membrane average thickness\cite{phillips2009,wiggins2005}, or
a non-cylindrical shape of the transmembrane region
\cite{phillips2009,wiggins2005,kim1998}.  Beside the hydrophobic
mismatch itself, the strength, and even the sign of such interactions
depend on many factors, including membrane stiffness to bending and
stretching, and the monolayer's spontaneous curvature.  The range of
these interactions is comparable to the protein size itself, and hence
could be expected to influence the effect of crowding on the gating
energy significantly.

\begin{figure}[!t]
\begin{center}
  \includegraphics{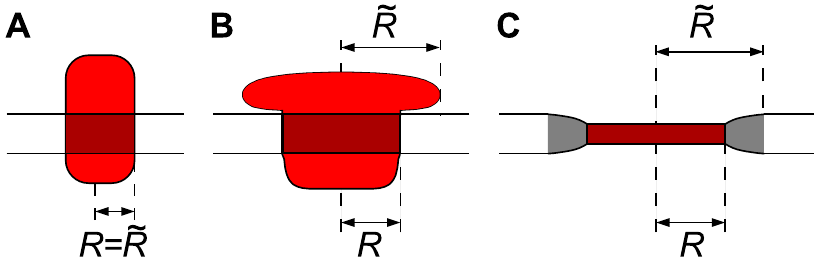}
\end{center}
  \caption{{\bf Mechanisms for different excluded
        area for proteins and lipids.}  One difference between the
      hard disk model of membrane proteins (A), and
      more complex protein structures (B,C) might be
      thought of in terms of different effective radii $\tilde R$ and
      $R$ for steric exclusion of surrounding proteins and lipids
      respectively. A protein (red) with a large domain outside of the
      bilayer (B) might exclude surrounding proteins, but not lipids,
      from approaching the transmembrane domain (dark red). Similarly,
      proteins with different hydrophobic thickness than the
      surrounding bilayer (C) generate a local zone of deformed lipid
      bilayer (gray) that effectively excludes other well-matched
      proteins. Horizontal lines indicate the surrounding lipid
      bilayer.}\label{overhang}
\end{figure}

The rich and interesting many body effects that can emerge from local
membrane deformations are outside the scope of this paper. However,
our calculations offers some qualitative insight into the sensitivity
of the crowding effect to structural features of the involved
proteins, which also includes some effects of hydrophobic mismatch.
In our hard disk calculations under constant tension, the entropic
surface tension cancels from the gating energy contribution (between
Eqs.~\eqref{crowdingtensions} and \eqref{crowdingtensionsCT}\ML{, and
  in \Eq{dGtension1}}), when the increase in channel area is balanced
by an increased total area. This cancellation reflects an underlying
assumption in the disk model of membrane proteins, which effectively
models membrane proteins as cylinders (figure \ref{overhang}A), from
which lipids and other proteins experience the same area exclusion.

  Real membrane proteins, however, can have complex shapes that
  violate this assumption \cite{CR:engelman2005}, for example due to
  large domains outside the bilayer that do not directly affect the
  local ordering of the lipids, but provide additional steric
  interaction with other membrane proteins, as sketched in figure
  \ref{overhang}B.  \ML{Many membrane bound receptors that bind bulky
    ligands near the membrane surface
    \cite{GPCR:rasmussen2011,hGH:clackson1995,farrens2010,sharif-naeini2010},
    yield complexes with a similar shape.  We would expect significant
    crowding effects on both the binding kinetics and the stability of
    the complex for these systems, similar to what has been seen for
    surface
    adsoption\cite{CR:chatelier1996,CR:minton1999,CR:minton2000,CR:leventis2010}.
    There are also examples of membrane proteins whose bulky
    cytoplasmic domains undergo substantial conformational changes,
    such as the mechanosensitive channel
    MscS\cite{MscS:grajkowski2005,MscS:nomura2008} and the
    $\mathrm{Ca}^{2+}\text{ATPase}$\cite{Ca:olesen2007}.} Hydrophobic
  mismatch \ML{might} play a similar role in a surrounding of mostly
  well-matched proteins (figure \ref{overhang}C).

  The presence of conformations with such structural features might
  remove the surface tension cancellations, and thereby change the
  dependence of gating energy on crowding in a qualitative way.  One
  should thus consider the two ensembles studied here (constant area
  and constant tension) as limiting cases capturing the range of
  phenomenon that can be expected for real membrane proteins.
  \ML{Trying to imagine more quantitative estimates of these effects
    points towards new and interesting questions, both theoretically
    and regarding structural features of whole membrane proteomes.
    For example, it seems likely that the large cytoplasmic domain of
    MscS experiences a different crowding environment than it's
    transmembrane part.  First, a large cytoplasmic domain can be
    crowded by macromolecules in solution\cite{MscS:grajkowski2005}.
    Second, it can only interact directly with those membrane proteins
    that also possess bulky cytoplasmic domains, not with those that
    mainly consists of transmembrane helices. Finally, the MscS
    transmembrane part might be shielded from direct interaction with
    the transmembrane parts of other proteins with bulky cytoplasmic
    domains, if those domains are large enough.}

The present analysis has as its key outcome the hypothesis that under
sufficiently crowded conditions, membrane proteins can influence each
others conformational changes through an entropic tension.  Though we
explored the consequences of that idea for one particular channel,
given the great diversity of membrane proteins and the high degree of
crowding in many membrane types, we expect that such effects could be
common.

\pdfbookmark[1]{Acknowledgments}{Acknowledgments}
\section*{Acknowledgments}
We are grateful to KC Huang, Jane Kondev, Doug Rees, Matthew Turner,
Tristan Ursell and Paul Wiggins for insightful discussions.

This study was funded by the National Institutes of Health through NIH
Award number R01 GM084211 and the Directors Pioneer Award, grant DP1
OD000217A (R.P. and M.L.), www.nih.gov, La Fondation Pierre Gilles de
Gennes (R.P. and P.S.), www.fondation-pgg.org, the Wenner-Gren
foundations (M.L.), www.swgc.org, and the Foundations of the Royal
Swedish Academy of Sciences (M.L.), www.kva.se. The funders had no
role in study design, data collection and analysis, decision to
publish, or preparation of the manuscript.


\pdfbookmark[1]{References}{References}

\clearpage
\setcounter{section}{0}
\setcounter{subsection}{0}
\setcounter{subsubsection}{0}
\setcounter{equation}{0}
\renewcommand{\thesection}{S\arabic{section}}
\renewcommand{\theequation}{S.\arabic{equation}}

\begin{flushleft}
\pdfbookmark[1]{Supporting text S1}{SI}
{\Large
\textbf{Entropic Tension in Crowded Membranes -- Supporting text S1}
}
\\
Martin Lind\'en$^{1,2}$, 
Pierre Sens$^{3}$, 
Rob Phillips$^{1,3,4,\ast}$
\\
\bf{1} Dept. of Applied Physics, California Institute of Technology, Pasadena, California, U.S.A
\\
\bf{2} Present address: Dept. of Biochemistry and Biophysics, Stockholm University, Stockholm, Sweden
\\
\bf{3} Laboratoire de Physico-Chimie Th{\'e}orique CNRS/UMR 7083 - ESPCI, 75231 Paris Cedex 05, France

\bf{4} Division of Biology, California Institute of Technology, Pasadena, California, U.S.A
\\
$\ast$ E-mail: phillips@pboc.caltech.edu
\end{flushleft}

\pdfbookmark[1]{Neglecting lipid elasticity}{SI:1}
\section*{Neglecting lipid elasticity}\label{SI:squishing}
Our calculations in the constant area ensemble neglected changes in
the elastic energy of the lipid bilayer. Similarly, we approximated
the constant tension ensemble with a constant lipid area, i.e., we
assumed that the strain in the lipids does not change significantly
upon gating. We also neglected thermal fluctuations in lipid area.

In this and the following section, we motivate these assumptions,
by using a simple model of lipid elasticity to argue that these
effects only gives rise to small corrections, so that neglecting them
is a legitimate approximation.  We will use the harmonic model for a
membrane patch with area $A$ and unstretched lipid area $A_0$,
\begin{equation}\label{Lenergy}
  H_\text{lipids}=\frac{\kappa_A}{2}A_0\left(\frac{A-n\pi\mean{R^2}-A_0}{A_0}\right)^2
  =\frac{\kappa_A}{2A_0}(A-A_{\vec{n}})^2,
\end{equation}
where we used the definition $A_{\vec{n}}=A_0+n\pi\mean{R^2}$ in the
last equality. Many lipid bilayers have an area compression modulus
$\kappa_A$ around $\text{0.25 J/m}^2 \sim 60\,\kBT/\nm^2$ (after
correction for bending fluctuations), with cholesterol-rich and
biological membranes being even stiffer \cite{boal}.


If we assume no prestretching of the lipid bilayer, so that
$(A-A_{\vec{n}})\sim \Delta A$, then the change in lipid elastic
energy (\Eq{Lenergy}) is of the order $\Delta H_\text{lipids}\sim
\kappa_A\Delta A^2/2A_0$. With our numerical test parameters
($A_0=A(1-\phi)=A/2\text{, }\Delta A=5\pi\text{ nm}^2\text{,
}\kappa_A\approx 60 \kBT/\nm^2$), and a patch area on the order
$1\;\mu\text{m}^2$, this gives $\Delta H_\text{lipids}\sim 0.02 \kBT$,
which can be safely neglected.
\pdfbookmark[1]{Constant lipid area approximation}{SI:2}
\section*{Constant lipid area approximation}\label{SI:constarea}
To motivate the approximation of constant lipid area in the constant
tension ensemble, we use the above lipid elastic model to estimate the
contributions from lipid elasticity and area fluctuations to the
gating energy.  We start with some numerical estimates.

Typical MscL gating tensions are $\sigma_0\sim 0.3-1.3\;\kBT/\nm^2$,
which gives rise to a lipid strain of
$\frac{\sigma_0}{\kappa_A}\approx 0.005-0.02$. The crowding surface
tension with our numerical test parameters ($\phi=0.5$ protein area
fraction, $R_p=1\;\nm$, $c_A=0.16\text{ protein/nm}^2$, and
$\eta^2=0$), is $-\sigma_\text{SPT}=\kBT
c_A/(1-\phi)^2=0.64~\kBT/\nm^2$, which induces a strain of magnitude
$10^{-2}$, comparable the gating strain.

The crowding tension changes with total area, in a way that can be
characterized by an entropic compression modulus,
\begin{equation}\label{kSPT1}  
  \kappa_\text{SPT}=-A_0\kBT\frac{\partial^2\ln Z}{\partial A^2}
  =A_0\frac{\partial\sigma_\text{SPT}}{\partial A}.
\end{equation}
Differentiating the expression for $\sigma_\text{SPT}$,
\Eq{SPTtension1} in the main text, we get
\begin{equation}\label{kSPT2}      
      \kappa_\text{SPT}
      =\frac{nA_0\kBT}{\big(A-n\pi\mean{R^2}\big)^{2}}
      +\frac{2A_0\kBT n^2\pi\mean{R}^2}{\big(A-n\pi\mean{R^2}\big)^{3}}.
\end{equation}
In this expression, $A_0$ is truly a constant (by definition of the
compression modulus), but since we expect the lipid strain to be
small, we can get a numerical estimate by using $A_0\approx
A-n\pi\mean{R^2}=A(1-\phi)$, which allows us to simplify in the usual
way, and write
\begin{equation}\label{kSPT3}
   \kappa_\text{SPT}\approx
   \frac{c_A\kBT}{(1-\phi)}\Big(1+\frac{c_A2\pi\mean{R}^2}{1-\phi}\Big).
\end{equation}
With our numerical test parameters and $\kappa_A=60 ~\kBT/\nm^2$, we
get $\kappa_\text{SPT}/\kappa_A\approx 0.016$, so the entropic
compression modulus is insignificant compared to the lipid stiffness.

In light of the above tension estimates, an estimated upper bound on
the lipid strain near MscL gating conditions is therefore
    \begin{equation}
      \frac{\sigma_0-\sigma_\text{SPT}}{\kappa_A+\kappa_\text{SPT}}
      \lesssim \frac{2\sigma_0}{\kappa_A}\sim 0.03.
    \end{equation}
This is not quite as small as $\pi R^2/A$, but still small enough that
we can safely neglect terms of that magnitude.
 
Next, we assess the importance of lipid area changes and fluctuations
by computing the partition function $\Xi$ using a saddle-point
approximation, i.e., expand fluctuations to second order, and show
that we recover the results of the main text, plus small correction
terms.

The above estimates strongly suggest that only areas close to
$A_{\vec{n}}$ will contribute significantly to the constant tension
partition function $\Xi$ (see \Eq{CTpartition} in the main text), and we
therefore expand the canonical partition function $Z$ to second order
around that point. We get
\begin{equation}
  \ln Z(A,\vec{n},T)=
  \ln Z(A_{\vec{n}},\vec{n},T)  
  +\Big.\frac{\partial\ln Z}{\partial A}\Big|_{A_{\vec{n}}}(A-A_{\vec{n}})
  +\frac{1}{2}\frac{\partial^2\ln Z}{\partial A^2}\Big|_{A_{\vec{n}}}
  (A-A_{\vec{n}})^2+\ldots.
\end{equation} 
Upon substituting \Eq{kSPT1} and \Eq{SPTtension0} in the main text, we can
write this in the form
\begin{equation}
  \ln Z(A,\vec{n},T)\approx
  \ln Z(A_{\vec{n}},\vec{n},T)-\beta\sigma_\text{SPT}(A-A_{\vec{n}})\\
  -\frac{\beta\kappa_\text{SPT}}{2A_0}(A-A_{\vec{n}})^2,
\end{equation}
where it is understood that $\sigma_\text{SPT}$ and
$\kappa_\text{SPT}$ are evaluated at $A=A_{\vec{n}}$.  Inserting the
simple lipid energy and the above expansion in the partition function
$\Xi$ (main text \Eq{CTpartition}), we get
  \begin{align}
  \Xi(\sigma,\vec{n},T)=&\int dA\;Z(A,\vec{n},T) 
  \exp\left(\beta\sigma A-\frac{\beta\kappa_A}{2A_0}(A-A_{\vec{n}})^2\right)\\
  \label{Xigauss}
  \approx&
  Z(A_{\vec{n}},\vec{n},T)e^{\beta\sigma A_{\vec{n}}}
  \int dA\; \exp\left(\beta(\sigma-\sigma_\text{SPT})(A-A_{\vec{n}})
  -\frac{\beta(\kappa_A+\kappa_\text{SPT})}{2A_0}(A-A_{\vec{n}})^2\right).
  \end{align}
The function in the exponent takes a maximum at
$A^*_{\vec{n}}=A_{\vec{n}}+A_0\fix{\frac{\sigma-\sigma_\text{SPT}}{\kappa_A+\kappa_\text{SPT}}}_{A_{\vec{n}}}$,
which is indeed very close to $A_{\vec{n}}$, since $A_0<A_{\vec{n}}$,
and $\frac{\sigma-\sigma_\text{SPT}}{\kappa_A+\kappa_\text{SPT}}\ll
1$. We can now evaluate $\Xi$ approximately, since \Eq{Xigauss} is a
Gaussian integral, and get the saddle point approximation
\begin{align}
  \Xi(\sigma,\vec{n},T)\approx&
  Z(A_{\vec{n}},\vec{n},T)e^{\beta\sigma A_{\vec{n}}}
  \exp\left(\frac{\beta A_0}{2}
  \frac{(\sigma-\sigma_\text{SPT})^2}{\kappa_A+\kappa_\text{SPT}}
\right)\fix{\sqrt{\frac{2\pi\kBT A_0}{\kappa_A+\kappa_\text{SPT}}}}_{A_{\vec{n}}}.
\end{align}
After discarding irrelevant constants, we recover the free energy of
\Eq{XiLegendre} in the main text, with two correction terms that we denote
$G_1$ and $G_2$,
\begin{equation}\label{XiExpansion2}
  \underbrace{G(\sigma,\vec{n},T)\approx -\kBT\ln
    Z(A_{\vec{n}},\vec{n},T) -\sigma A_{\vec{n}}}_{\text{\Eq{XiLegendre}}} 
  \underbrace{-\frac{A_0}{2}
    \frac{(\sigma-\sigma_\text{SPT})^2}{\kappa_A+\kappa_\text{SPT}}
  }_{=G_1(A_{\vec{n}},\vec{n},T)}
  +
  \underbrace{
    \frac{\kBT}{2}\ln\big(1+\frac{\kappa_\text{SPT}}{\kappa_A}\big)
    }_{=G_2(A_{\vec{n}},\vec{n},T)},
\end{equation}  
with all terms evaluated at area $A_{\vec{n}}$.  What we need to show,
is that the correction terms $G_1$ and $G_2$ indeed produce only small
contributions to the overall gating tension.

We start with $G_2$. Since $\frac{\kappa_\text{SPT}}{\kappa_A}\ll 1$,
we can Taylor expand the logarithm, and obtain $G_2\approx
\kBT\kappa_\text{SPT}/2\kappa_A$. But for the same reason, we see that
$G_2$ is already a small quantity, and therefore cannot make a
significant contribution to the gating free energy.

For $G_1$, a little more effort is needed. We start by noting that we
can ignore the entropic compression modulus, since it only contributes
a factor $\frac{\kappa_A}{\kappa_\text{SPT}+\kappa_A}\approx
(1-\frac{\kappa_\text{SPT}}{\kappa_A}+\ldots)$, i.e., a small
correction to $\Delta G_1$.  Next, we rewrite $\sigma_\text{SPT}$ from
\Eq{SPTtension1} in the main text, substitute $\Var(R)=\mean{R^2}-\mean{R}^2$,
and use the fact that $A-n\pi{R^2}=A_0$ when $A=A_{\vec{n}}$. This
leads to
\begin{equation}\label{sigmaG1}
  \frac{\sigma_\text{SPT}}{\kBT}
  =-n\frac{A_0+n\pi\mean{R}^2}{A_0^2}=
  -\frac{n}{A_0}-\Big(\frac{n}{A_0}\Big)^2\pi\mean{R}^2.
\end{equation}

We now consider the gating energy contribution, for the case of $N$
crowders and a single channel that changes radius from $R_c$ to
$R_o$. Using the definition of $G_1$ in \Eq{XiExpansion2}, we get
\begin{equation}
  \Delta G_1=-\frac{A_0}{2\kappa_A}\Big(
  \big(\sigma-\sigma_\text{SPT}(\vec{N}+\hat{e}_o)\big)^2
  -\big(\sigma-\sigma_\text{SPT}(\vec{N}+\hat{e}_c)\big)^2\Big),
\end{equation}
where we have written out the copy number distribution arguments of
the tensions. This can be rewritten as
\begin{equation}
    \Delta G_1=-\frac{A_0}{\kappa_A}\bigg(\sigma
  -\frac{\sigma_\text{SPT}(\vec{N}+\hat{e}_o)
        +\sigma_\text{SPT}(\vec{N}+\hat{e}_c)}{2}
  \bigg) \times\Big(
  \sigma_\text{SPT}(\vec{N}+\hat{e}_c)
 -\sigma_\text{SPT}(\vec{N}+\hat{e}_o)\Big).
\end{equation}
For order of magnitude estimates, approximating
$\frac{1}{2}\big(\sigma_\text{SPT}(\vec{N}+\hat{e}_o)
+\sigma_\text{SPT}(\vec{N}+\hat{e}_c)\big)\approx \sigma_\text{SPT}$
(with unspecified channel conformation) is good enough. The difference
in the last factor must be treated with more care. Substituting
\Eq{sigmaG1} with $n=N+1$ and
$\mean{R}=\frac{N\mean{R_p}+R_{o,c}}{N+1}$ for the mean protein radius
in the two states ($\mean{R_p}$ is the mean radius of the crowders
only), we get
\begin{equation}
  \frac{\Delta G_1}{\kBT}
  =-A_0\frac{\sigma-\sigma_\text{SPT}}{\kappa_A}
    \times\Big(\frac{N+1}{A_0}\Big)^2
\Big(\frac{(N\mean{R_p}+R_o)^2-(N\mean{R_p}+R_c)^2}{(N+1)^2}\Big).
\end{equation}
Expanding the squares in the last factor, substituting
$\frac{N}{A_0}=\frac{c_A}{1-\phi}$, and neglecting a term of order
$\pi R^2/A_0$, we finally get
\begin{equation}
  \frac{\Delta G_1}{\kBT}
  =-\Big(\frac{\sigma-\sigma_\text{SPT}}{\kappa_A}\Big)
  \frac{c_A\mean{R_p}\Delta C}{1-\phi}.
\end{equation}
This is a negative correction to the familiar line tension. For gating
of MscL ($\sigma\approx \sigma_0$), it is smaller than the lowest
order term, \Eq{dGresult} in the main text, by a factor
$\frac{\sigma_0-\sigma_\text{SPT}}{\kappa_A}\lesssim 10^{-2}$, indeed
a small correction.


\end{document}